\documentclass[
a4paper, % Paper size, use either a4paper or letterpaper
11pt,    % Default font size, can also use 11pt or 12pt, although this is not recommended
unnumberedsections, % Comment to enable section numbering
twoside, % Two side traditional mode where headers and footers change between odd and even pages, comment this option to make them fixed
]{LTJournalArticle}

\usepackage{amsmath}
\usepackage{amsfonts}
\usepackage{amssymb}
\usepackage{amstext}
\usepackage{amsthm}
\usepackage{gensymb}

\newtheorem{remark}{Remark}
\newcounter{RomanNumber}

\usepackage{algorithm}
\usepackage{algorithmic}
\usepackage{color}
\usepackage{makecell}
\runninghead{Shortened Running Article Title} % A shortened article title to appear in the running head, leave this command empty for no running head

\footertext{\textit{} (XXXX) X: XXX-XXX} % Text to appear in the footer, leave this command empty for no footer text

\title{Angle Estimation of a Single Source with Massive Uniform Circular Arrays} % Article title, use manual lines breaks (\\) to beautify the layout

\author{Mingyan Gong\textsuperscript{1}
\thanks{Corresponding author: \href{mailto:jinyan0\_o@outlook.com}{jinyan0\_o@outlook.com} \\ \textbf{Received:} XXXX XX, XXXX, \textbf{Published:} XXXX XX, XXXX}
}

\date{\footnotesize\textsuperscript{\textbf{1}} School of Electronics and Information Engineering, Huaibei Institute of Technology %\\ %\textsuperscript{\textbf{2}}Physics Department, 
}

\renewcommand{\maketitlehookd}{
	\begin{abstract}
		\noindent Estimating the directions of arrival (DOAs) of incoming plane waves is an essential topic in array signal processing. Widely adopted uniform linear arrays can only provide estimates of source azimuth. Thus, uniform circular arrays (UCAs) are attractive in that they can provide $360\degree$ azimuthal coverage and additional elevation angle information. Considering that with a massive UCA, its polar angles of array sensors can approximately represent azimuth angles over $360\degree$ using angle quantization, a simple two-dimensional DOA estimation method for a single source is proposed. In this method, the quantized azimuth angle estimate is obtained by only calculating and comparing a number of covariances, based on which the elevation angle estimate is then obtained by an explicit formula. Thus, the proposed method is computationally simple and suitable for real-time signal processing. Numerical results verify that the proposed method can obtain azimuth as well as elevation angle estimates and the estimates can be used as starting points of multidimensional searches for methods with higher accuracy. Additionally, the proposed method can still work in the presence of nonuniform noise.

        \par\textbf{Keywords: } Array signal processing, direction of arrival (DOA) estimation, massive multiple-input multiple-output (MIMO), nonuniform noise, two-dimensional (2–D) angle estimation, uniform circular array.%\par在段首，表示另起一行
	\end{abstract}
}

\begin{document}

\maketitle % Output the title section

\section{Introduction}
Uniform linear arrays (ULAs) are widely used in the direction of arrival (DOA) estimation literature due to the simplicity \cite{Krim}. However, ULAs fail to provide any information on source elevation angles. Toward this end, planar arrays are adopted when estimates of source azimuth and elevation are required. In particular, uniform circular arrays (UCAs) are attractive due to the advantages, e.g., $360\degree$ azimuthal coverage, additional elevation angle information, and almost invariant directional pattern.

It is well-known that high-resolution methods for two-dimensional (2-D) DOA estimation with UCAs include subspace and maximum-likelihood (ML) based methods, e.g., the UCA-RM-MUSIC and UCA-ESPRIT algorithms \cite{Mathews}, the modified 2-D MUSIC algorithm \cite{Ye}, the alternating maximization algorithm \cite{Ziskind}, and the expectation-maximization type algorithms \cite{Feder, Miller, Gong1, Gong2}. However, these algorithms need eigendecomposition and/or multidimensional search, which are computationally demanding. Moreover, the global matched filter based method in \cite{Fuchs} requires 2-D search, and is also unsuitable for real-time signal processing. To avoid the high complexity of eigendecomposition and/or multidimensional search, a 2-D angle estimation method for a single source with UCAs in \cite{Wu} is developed. This method is without eigendecomposition and multidimensional search since the angle estimates are obtained by explicit formulas. However, this method requires that the number of array sensors be even. To eliminate the requirement, a generalized method is further proposed in \cite{Liao}, which also needs neither eigendecomposition nor multidimensional search. In this paper, we propose a computationally simpler 2-D angle estimation method for a single source with massive UCAs.

As a very promising enabling technique, massive arrays or multiple-input multiple-output (MIMO) systems have received considerable attention \cite{Rusek}. Since the beam width of main lobe in an array becomes narrower and narrower as the number of array sensors increases, massive MIMO systems require the ultra-high accuracy of DOA estimation for performing downlink beamforming. Fortunately, by utilizing a large number of array sensors, massive arrays are able to achieve high angular resolution and accuracy. As a consequence, massive MIMO systems have been combined with other techniques to further improve localization accuracy or reduce computational complexity. In \cite{Cao}, a low-complex one-snapshot DOA estimation method with massive ULAs is proposed while in \cite{Swindlehurst}, the combination of massive MIMO and millimeter-wave techniques is considered.

Actually, the possible array configurations of massive MIMO systems include not only ULAs but also UCAs \cite{Li,Hu}. Considering that with a massive UCA, its polar angles of array sensors in polar coordinates can approximately represent azimuth angles over $360\degree$ using angle quantization, a simple method for 2-D angle estimation of a single source is proposed in this paper. In this method, the quantized azimuth angle estimate is obtained by only calculating and comparing a number of covariances, based on which the elevation angle estimate is then obtained by an explicit formula. Although the proposed method cannot yield sufficient accuracy due to angle quantization, it is computationally simple and suitable for real-time signal processing. Additionally, its estimates can be used as good starting points of multidimensional searches for methods with higher accuracy, e.g., ML direction finding based descent algorithms. Finally, numerical results verify the effectiveness of the proposed method.

\emph{Notations}: $\Re\{a\}$, $\Im\{a\}$, $\vert a\vert$, and $a^{\ast}$ mean the real part, imaginary part, modulus, and conjugate number of complex number $a$, respectively. $\mathbf{a}^T$ and $\Vert\mathbf{a}\Vert$ denote the transposition and Euclidean norm of vector $\mathbf{a}$, respectively. $\mathrm{E}\{\cdot\}$ stands for the expectation operator and $\mathrm{D}\{\cdot\}$ is the variance operator. $\jmath$ denotes the imaginary unit and $\lfloor\cdot\rfloor$ represents the floor operator.

\section{Signal Model}

\begin{figure}[t] \centering\vspace{0cm}
\includegraphics[scale=0.6]{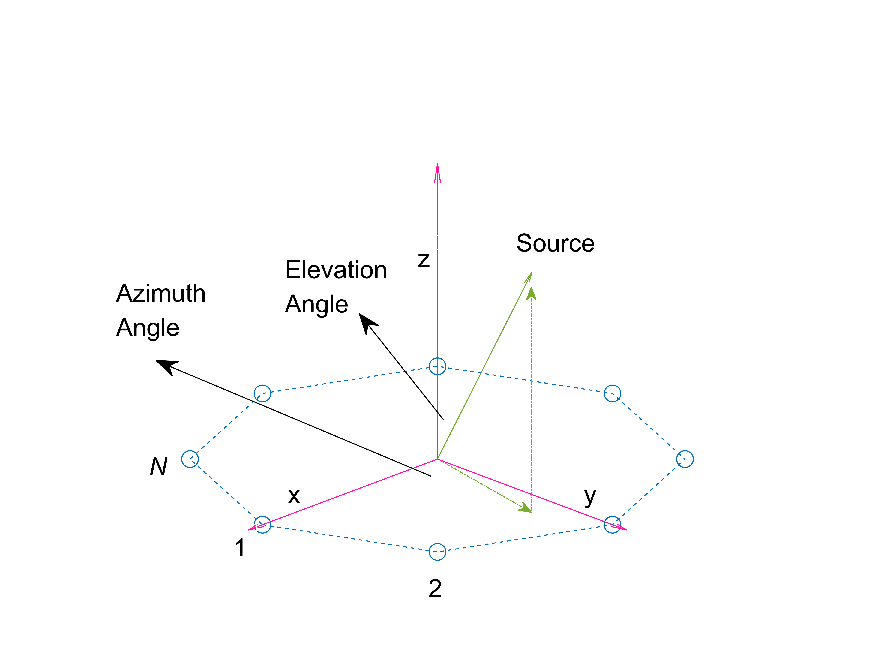}
\vspace{0cm}\caption{Signal model.}\label{fig1}\vspace{0cm}
\end{figure}

As depicted in Fig. \ref{fig1}, we consider a UCA with $N$ isotropic sensors, which are distributed over the circumference of a circle with radius $R$ in the $xy$ plane. The spherical coordinate system is utilized to show the DOA of a source impinging on this UCA. The origin is located at the center of this array for simplicity. As a result, the $n$-th array sensor is with polar angle $\theta_{n}=\frac{2\pi}{N}(n-1),n=1,2,\dots,N$.

Assume that only one narrow-band source of elevation angle $\varphi \in (0, \frac{\pi}{2})$ and azimuth angle $\phi\in [0, 2\pi)$ from far field impinges on this UCA. Hence, the phase difference between the source signals received at the origin and the $n$-th array sensor is $\psi_n = -\zeta\sin(\varphi)\cos(\phi-\theta_{n})$, where $\zeta = 2\pi\frac{R}{\lambda}$ and $\lambda$ is the wavelength of the source. Then, the composite signal received at the $n$-th sensor is expressed as
\begin{equation} \label{1}    %eq1
x_{n}(t)=e^{\jmath\psi_n}s(t)+w_{n}(t), n = 1,2,\ldots, N,
\end{equation}
where $s(t)$ is the source signal received at the origin and its power is $P$, i.e., $\mathrm{E}\{|s(t)|^2\}=P$. Moreover, $w_{n}(t)$ denotes a spatially uniform white Gaussian noise with zero mean and variance $\sigma^2$, i.e., $w_n(t)\sim\mathcal{CN}(0,\sigma^2)$. In \eqref{1}, the $s(t)$ and $w_{n}(t)$'s are assumed to be mutually uncorrelated.

\begin{remark}
In various practical scenarios, the spatially uniform white noise assumption may be violated. Spatially nonuniform white noise is present in an array when noise variances in different sensors are not identical, which makes the noise covariance matrix diagonal. To address the issue of performance degradation under nonuniform noise, many DOA estimation methods have been proposed \cite{Pesavento, Madurasinghe, Esfandiari, Gong3, Gong4, Stoica}.

It is worth pointing out that when the spatially nonuniform white noise assumption is adopted, the method proposed in this paper can be applied directly. For simplicity and illustration, we only use spatially uniform white noise.
\end{remark}

\subsection{Covariance Analysis}

We write the covariance between the two signals received at the $i$-th and $j$-th sensors as
\begin{eqnarray}   %eq2
r_{i,j}=\mathrm{E}\big\{x_{i}(t)x_{j}^{\ast}(t)\big\}=Pe^{\jmath(\psi_i - \psi_j)}, i\ne j.
\end{eqnarray}
Note that once $r_{i,j}$ is determined, its argument will be derived as
\begin{eqnarray}   %eq3
&~&\psi_i - \psi_j = \angle(r_{i,j}) \nonumber \\
&=& - \zeta\sin(\varphi)[\cos(\phi-\theta_{i})-\cos(\phi-\theta_{j})], 
\end{eqnarray}
 where $|\psi_i - \psi_j| < 2\zeta$. In order to guarantee that there is no phase ambiguity in (3), let $2\zeta\le\pi$, i.e., $\frac{R}{\lambda} \le \frac{1}{4}$, $-\pi < \psi_i - \psi_j < \pi$, and
\begin{equation}   %eq4
\psi_i - \psi_j =
\left\{
\begin{array}{ll}
{\Psi_{i,j},} & {\Re\{r_{i,j}\} > 0}, \\
{\Psi_{i,j}-\pi,} & {\Re\{r_{i,j}\} < 0,\Im\{r_{i,j}\} < 0}, \\
{\Psi_{i,j}+\pi,} & {\Re\{r_{i,j}\} < 0,\Im\{r_{i,j}\} > 0}, \\
{\frac{\pi}{2},} & {\Re\{r_{i,j}\} = 0,\Im\{r_{i,j}\} > 0},  \\
{-\frac{\pi}{2},} & {\Re\{r_{i,j}\} = 0,\Im\{r_{i,j}\} < 0},
\end{array}
\right.
\end{equation}
where $\Psi_{i,j} = \arctan\big(\frac{\Im\{r_{i,j}\}}{\Re\{r_{i,j}\}}\big)$.

From the perspective of identifiability of any parametrization \cite{Stoica}, when two unknowns (free parameters) exist, there are least two equations (constraints) to determine both unknowns. According to this principle, some simple methods based on the least squares estimation have been proposed in \cite{Wu,Liao}. In the next section, we will design a computationally simpler method.

\subsection{Method Principle}

Recalling (3) and (4), we can see that deriving $\psi_i-\psi_j$ usually requires the operation of the inverse trigonometric function $\arctan\big(\frac{\Im\{r_{i,j}\}}{\Re\{r_{i,j}\}}\big)$, but $\psi_i-\psi_j = 0$ is the \emph{only} argument obtained directly without calculations due to $r_{i,j}$ being a real number. This shines a light on how to satisfy $\psi_i - \psi_j = 0$. To proceed, we rewrite (3) as
\begin{equation} \label{5}  %eq5
\psi_i - \psi_j = -2\zeta\sin(\varphi)\sin(\phi-\vartheta_{i,j})\sin(\varpi_{i,j}), i\ne j,
\end{equation}
where $\vartheta_{i,j}=\frac{\theta_{i}+\theta_{j}}{2}$, $\varpi_{i,j}=\frac{\theta_{i}-\theta_{j}}{2}\in (-\pi, 0)\cup(0, \pi)$, and $\phi - \vartheta_{i,j}\in (-2\pi, 2\pi)$. Then, $\psi_i - \psi_j = 0$ if and only if
\begin{eqnarray}
\phi\in\{\vartheta_{i,j}, \vartheta_{i,j}+\pi\} ~&\text{when}&~\vartheta_{i,j} \in (0, \pi) \nonumber\\
\text{or}~\phi\in\{\vartheta_{i,j}, \vartheta_{i,j}-\pi\}~&\text{when}&~\vartheta_{i,j}\in[\pi, 2\pi).\nonumber
\end{eqnarray}

Unfortunately, $\vartheta_{i,j}$ is a discrete variable while $\phi$ is a continuous variable, so it is very difficult to utilize $\vartheta_{i,j}$ for estimating $\phi$ \emph{without bias}. On the other hand, 
\begin{itemize}
	\item it is normal that there is always an error in an estimator,
	\item with the number of array sensors $N$ increasing, i.e., a massive UCA is adopted, the value range of $\vartheta_{i,j}$ enlarges.
\end{itemize}
Thus, $\vartheta_{i,j}$ can be used to estimate $\phi$ in the form of quantization when $N$ is large.

%\begin{enumerate}
%	\item Arcu
%	\item Fusce
%\end{enumerate}

\section{Algorithm Design}

\subsection{Angle Quantization}
As mentioned above, we can use $\vartheta_{i,j}$ to estimate $\phi$ based on the fact that $r_{i,j}$ is a real number if $\vartheta_{i,j}=\phi$, $\phi - \pi$, or $\phi +\pi$ in \eqref{5}, which leads to $\Im\{r_{i,j}\}=0$. In practice, $r_{i,j}$ is estimated only by calculating statistically independent snapshots as follows:
\begin{equation}  \label{6}  %6
\hat{r}_{i,j}=\frac{1}{L}\sum_{l=1}^{L}x_{i}(l)x_{j}^{\ast}(l),
\end{equation}
where $L$ is the number of snapshots. According to the principle of quantization and \eqref{5}, the estimator (one of $\hat{\phi}$, $\hat{\phi}-\pi$, and $\hat{\phi}+\pi$) should be obtained by finding $\hat{r}_{i^{\ast},j^{\ast}}$ closest to a real number, i.e., $\Im\{\hat{r}_{i^{\ast},j^{\ast}}\}$ is closest to zero. Thus, we design the problem:
\begin{eqnarray} \label{7} %7
(i^{\ast}, j^{\ast}) = \arg \underset{1 \le i, j \le N, i\neq j}{\min}|\Im\{\hat{r}_{i,j}\}|,
\end{eqnarray}
which indicates that the solution requires an almost exhaustive search among all array sensors. 

Note that in \eqref{5}
\begin{eqnarray}
|\varpi_{i,j}|\to0\Rightarrow|\sin(\varpi_{i,j})|\to0
\Rightarrow|\psi_i-\psi_j|\to0\nonumber,
\end{eqnarray}
which interferes with finding the solution in \eqref{7}. To avoid this issue, we constrain the ordinals $i$ and $j$ through analyzing the estimator $\hat{Y}=\sin(\hat{\varphi})\sin(\hat{\phi}-\vartheta_{i,j})$. Obviously, $\hat{Y}$ satisfies
\begin{equation}  %8
\hat{Y} = \frac{1}{\sin(\varpi_{i,j})}\times\Big(-\frac{\angle(\hat{r}_{i,j})}{2\zeta}\Big),
\end{equation}
and
\begin{equation}  %9
\mathrm{D}\{\hat{Y}\} = \frac{1}{\sin^{2}(\varpi_{i,j})}\times\text{D}\Big\{-\frac{\angle(\hat{r}_{i,j})}{2\zeta}\Big\}.
\end{equation}
In order to minimize $\mathrm{D}\{\hat{Y}\}$ with respect to $\varpi_{i,j}$, we should maximize $|\sin(\varpi_{i,j})|$ and thus $\varpi_{i,j} = \pm\frac{\pi}{2}$, i.e., $|\theta_{i}-\theta_{j}| = \pi$.

For satisfying $|\theta_{i}-\theta_{j}| = \pi$ and simplicity, let $N = 4m$, where $m\gg1$ is a positive integer, $1\le i\le\frac{N}{2}$ ($0\le\theta_i<\pi$), and $j = i + \frac{N}{2}$ ($\theta_j = \theta_i + \pi$). As a result, problem \eqref{7} is simplified to
\begin{equation}   \label{10}   %10
i^{\ast} = \arg\underset{1 \le i \le \frac{N}{2}, j = i + \frac{N}{2}}{\min}|\Im\{\hat{r}_{i,j}\}|,
\end{equation}
which means that the optimal solution only requires $\frac{N}{2}$ covariances. After obtaining $i^{\ast}$, we use \eqref{5} to obtain
\begin{equation}       %11
\psi_{i^\ast}-\psi_{i^\ast+\frac{N}{2}}\approx 0\Rightarrow
\cos(\phi-\theta_{i^\ast})\approx 0, \phi\in[0,2\pi),
\end{equation}
where $\phi$ can be estimated as
\begin{equation}          %equation 12
\hat{\phi} =
\left\{
\begin{array}{ll}
{\theta_{i^\ast}+\frac{\pi}{2}~\text{or}~\theta_{i^\ast}+\frac{3}{2}\pi,} & {1\le i^\ast\le\frac{N}{4}}, \\
{\theta_{i^\ast}+\frac{\pi}{2}~\text{or}~\theta_{i^\ast}-\frac{\pi}{2},} & {\frac{N}{4}<i^\ast\le\frac{N}{2}}.
\end{array}
\right.
\end{equation}

To solve the phase ambiguity in (12), we utilize the two particular covariances $\hat{r}_{(i^{\ast}+\frac{N}{4}),(i^{\ast}+\frac{3}{4}N)}$ for $1\le i^{\ast}\le \frac{N}{4}$ and $\hat{r}_{(i^{\ast}-\frac{N}{4}),(i^{\ast}+\frac{N}{4})}$ for $\frac{N}{4}< i^{\ast}\le \frac{N}{2}$. Using both covariances, we have
\begin{subequations} \label{13}
\begin{eqnarray} \label{13a}  %13(a)
\psi_{(i^{\ast}+\frac{N}{4})} - \psi_{(i^{\ast}+\frac{3}{4}N)} \approx -2\zeta\sin(\varphi)\sin(\hat{\phi}-\theta_{i^\ast})\nonumber \\
=\left\{
\begin{array}{ll}
{-2\zeta\sin(\varphi)\in(-\pi, 0),} & {\hat{\phi}=\theta_{i^\ast} + \frac{\pi}{2}}, \\
{2\zeta\sin(\varphi)\in(0,\pi),} & {\hat{\phi}=\theta_{i^\ast} + \frac{3}{2}\pi},
\end{array}
\right.
\end{eqnarray}
\begin{eqnarray}     %13(b)
\psi_{(i^{\ast}-\frac{N}{4})} - \psi_{(i^{\ast}+\frac{N}{4})} \approx 2\zeta \sin(\varphi)\sin(\hat{\phi}-\theta_{i^\ast})\nonumber \\
 =\left\{
\begin{array}{ll}
{2\zeta \sin(\varphi)\in(0, \pi),} & {\hat{\phi}=\theta_{i^\ast} + \frac{\pi}{2}}, \\
{-2\zeta \sin(\varphi)\in(-\pi,0),} & {\hat{\phi}=\theta_{i^\ast} - \frac{\pi}{2}},
\end{array}
\right.
\end{eqnarray}
\end{subequations}
which indicates that the sign of $\Im\{\hat{r}_{(i^{\ast}+\frac{N}{4}),(i^{\ast}+\frac{3}{4}N)}\}$ or $\Im\{\hat{r}_{(i^{\ast}-\frac{N}{4}),(i^{\ast}+\frac{N}{4})}\}$ can be used to determine $\hat{\phi}$.
Specifically, when $1\le i^{\ast}\le\frac{N}{4}$, i.e., $\theta_{i^{\ast}}\in[0,\frac{\pi}{2})$,
\begin{eqnarray}
\Im\{\hat{r}_{(i^{\ast}+\frac{N}{4}),(i^{\ast}+\frac{3}{4}N)}\}<0 &\text{~for~}& \hat{\phi}=\theta_{i^\ast}+\frac{\pi}{2}, \nonumber \\ 
\Im\{\hat{r}_{(i^{\ast}+\frac{N}{4}),(i^{\ast}+\frac{3}{4}N)}\}>0 &\text{~for~}& \hat{\phi}=\theta_{i^\ast}+\frac{3}{2}\pi.\nonumber
\end{eqnarray}
When $\frac{N}{4}<i^{\ast}\le\frac{N}{2}$, i.e., $\theta_{i^{\ast}}\in[\frac{\pi}{2},\pi)$,
\begin{eqnarray}
\Im\{\hat{r}_{(i^{\ast}-\frac{N}{4}),(i^{\ast}+\frac{N}{4})}\}>0 &\text{~for~}& \hat{\phi}=\theta_{i^\ast} + \frac{\pi}{2}, \nonumber\\
\Im\{\hat{r}_{(i^{\ast}-\frac{N}{4}),(i^{\ast}+\frac{N}{4})}\}<0 &\text{~for~}& \hat{\phi}=\theta_{i^\ast}-\frac{\pi}{2}. \nonumber
\end{eqnarray}

Finally, from \eqref{13} $\varphi$ is estimated by
\begin{subequations}  \label{14}     %14
\begin{equation}     %14(a)
 \hat{\varphi} =
\left\{
\begin{array}{ll}
{\arcsin\big(\frac{\angle(\hat{r}_{(i^{\ast}+\frac{N}{4}),(i^{\ast}+\frac{3}{4}N)})}{-2\zeta}\big),} & {\hat{\phi}=\theta_{i^{\ast}}+\frac{\pi}{2}}, \\
{\arcsin\big(\frac{\angle(\hat{r}_{(i^{\ast}+\frac{N}{4}),(i^{\ast}+\frac{3}{4}N)})}{2\zeta}\big),} & {\hat{\phi}=\theta_{i^{\ast}}+\frac{3}{2}\pi}
\end{array}
\right.
\end{equation}
\begin{equation}     %14(b)
 \hat{\varphi} =
\left\{
\begin{array}{ll}
{\arcsin\big(\frac{\angle(\hat{r}_{(i^{\ast}-\frac{N}{4}),(i^{\ast}+\frac{N}{4})})}{2\zeta}\big),} & {\hat{\phi}=\theta_{i^{\ast}}+\frac{\pi}{2}},\\
{\arcsin\big(\frac{\angle(\hat{r}_{(i^{\ast}-\frac{N}{4}),(i^{\ast}+\frac{N}{4})})}{-2\zeta}\big),} & {\hat{\phi}=\theta_{i^{\ast}}-\frac{\pi}{2}}.
\end{array}
\right.
\end{equation}
\end{subequations}
The details of the proposed method are given in \textbf{Algorithm \ref{a1}}.

\begin{algorithm}
\caption{Angle Quantization Based 2-D Angle Estimation} \label{a1}
\begin{algorithmic}[1]
\STATE {Calculate $r(1)=\hat{r}_{1,1+\frac{N}{2}}$ in \eqref{6}.}
\STATE {Initialize $i^{\ast} = 1$, $min$ = $|\Im\{r(1)\}|$, $i = 2$.}
\WHILE{$i \le \frac{N}{2}$}
    \STATE {Initialize $j=i+\frac{N}{2}$ and calculate $r(i)=\hat{r}_{i,j}$ in \eqref{6}.}
    \IF{$|\Im\{r(i)\}|< min$}
        \STATE {Set $i^{\ast}=i$ and $min=|\Im\{r(i)\}|$.}
    \ENDIF
    \STATE {$i=i+1$.}
\ENDWHILE
\IF {$i^{\ast}\le\frac{N}{4}$}
       \IF {$\Im\{r(i^{\ast}+\frac{N}{4})\}<0$}
          \STATE {$\hat{\phi}=\theta_{i^{\ast}}+\frac{\pi}{2}$ and calculate $\hat{\varphi}$ in \eqref{14}.}
       \ELSE
          \STATE {$\hat{\phi}=\theta_{i^{\ast}}+\frac{3}{2}\pi$ and calculate $\hat{\varphi}$ in \eqref{14}.}
       \ENDIF
\ELSE
       \IF{$\Im\{r(i^{\ast}-\frac{N}{4})\}>0$}
          \STATE {$\hat{\phi}=\theta_{i^{\ast}}+\frac{\pi}{2}$ and calculate $\hat{\varphi}$ in \eqref{14}.}
       \ELSE
          \STATE {$\hat{\phi}=\theta_{i^{\ast}}-\frac{\pi}{2}$ and calculate $\hat{\varphi}$ in \eqref{14}.}
       \ENDIF
\ENDIF
\STATE {Output $\hat{\phi}$ and $\hat{\varphi}$.}
\end{algorithmic}
\end{algorithm}

\begin{remark}
Because of angle quantization, the proposed method cannot yield unbiased estimators and its estimation performance is inferior to that in \cite{Wu}, which is shown in Fig. \ref{f2}. However, the computational complexity of the proposed method is to search $i^{\ast}$ in \eqref{10} and the operation of the inverse trigonometric function in \eqref{14}, which makes it computationally simpler than the method in \cite{Wu} and suitable for real-time signal processing.

In addition, we can use the estimates provided by the proposed method as starting points of multidimensional searches for methods with higher accuracy. To this end, we will design an ML direction finding based descent algorithm using a few array sensors in the next subsection.
\end{remark}

\subsection{Maximum Likelihood Direction Finding}

It is well-known that in DOA estimation, ML based methods can yield sufficient accuracy \cite{Krim}. However, the ML based methods typically require multidimensional searches. Moreover, the ML based DOA estimation problems are nonconvex optimization problems and must be given good starting points before designing various descent algorithms. Obtaining good starting points, however, tends to be computationally intensive. Fortunately, the proposed method can be considered as a choice for good starting points.

We adopt the deterministic ML model, where the source signal waveform $s(t)$ is deterministic but unknown \cite{Feder,Miller,Gong1}. A subset of array sensors $\delta=\{i_1,\ldots,i_K\}$ with $K\ll N$ is chosen to reduce the computational complexity. Accordingly, $\hat{\phi}$ and $\hat{\varphi}$ can be obtained by solving the following minimization problem \cite{Ziskind}:
\begin{equation}  \label{15}     %15
\min_{\phi\in[0,2\pi),\varphi\in(0,\frac{\pi}{2})} -\mathrm{Tr}\{\mathbf{A}\hat{\mathbf{R}}\},
\end{equation}
where $\mathrm{Tr}\{\cdot\}$ denotes the trace operator, $\mathbf{A}$ and $\hat{\mathbf{R}}$ are two $K\times K$ Hermite matrices, whose elements of the $p$-th row and the $q$-th column are $a_{p,q}=e^{\jmath(\psi_{i_p}-\psi_{i_q})}$ and $\hat{r}_{i_p,i_q}$ in \eqref{6}, respectively.

The gradient descent method is used to solve problem \eqref{15} \cite{Boyd}. Let $l(\phi,\varphi)=-\mathrm{Tr}\{\textbf{A}\hat{\textbf{R}}\}$ and we first derive the gradient $\triangledown l(\phi,\varphi)=[\frac{\partial l}{\partial\phi}~\frac{\partial l}{\partial\varphi}]^T$ by
\begin{subequations} \label{16}
\begin{eqnarray}     %16(a)
\frac{\partial l}{\partial\phi}=&-2\zeta\sin(\varphi)\sum_{p=1}^{K-1}\sum_{q=p+1}^{K}[-\sin(\phi-\theta_{i_p})\nonumber\\
&+\sin(\phi-\theta_{i_q})]\Im\{a_{p,q}\hat{r}_{i_q,i_p}\},
\end{eqnarray}
\begin{eqnarray}     %16(b)
\frac{\partial l}{\partial\varphi}=&-2\zeta\cos(\varphi)\sum_{p=1}^{K-1}\sum_{q=p+1}^{K}[\cos(\phi-\theta_{i_p})\nonumber\\
&-\cos(\phi-\theta_{i_q})]\Im\{a_{p,q}\hat{r}_{i_q,i_p}\}.
\end{eqnarray}
\end{subequations}
This leads to the minimizing sequence at the $k$-th iteration as
\begin{subequations}
\begin{eqnarray} \label{17a}    %17a
\phi^{(k+1)}=\phi^{(k)}-t^{(k)}\Big(\frac{\partial l}{\partial\phi}\Big)^{(k)}\in(-\infty,\infty),
\end{eqnarray}
\begin{eqnarray}     %17b
\varphi^{(k+1)}=\varphi^{(k)}-t^{(k)}\Big(\frac{\partial l}{\partial\varphi}\Big)^{(k)}\in(0,\frac{\pi}{2}).
\end{eqnarray}
\end{subequations}
In \eqref{17a}, $\phi$ is not restricted to $[0,2\pi)$ and the step size $t\ge0$ satisfies
\begin{equation} \label{18}      %18
 t^{(k)} < t_0^{(k)}=
\left\{
\begin{array}{ll}
-\frac{\frac{\pi}{2}-\varphi^{(k)}}{\big(\frac{\partial l}{\partial\varphi}\big)^{(k)}},&\big(\frac{\partial l}{\partial\varphi}\big)^{(k)}<0,\\
\frac{\varphi^{(k)}}{\big(\frac{\partial l}{\partial\varphi}\big)^{(k)}},&\big(\frac{\partial l}{\partial\varphi}\big)^{(k)}>0.
\end{array}
\right.
\end{equation}
Furthermore, the backtracking line search is used to obtain $t^{(k)}$ and the details are given in \textbf{Algorithm \ref{a2}}.

\begin{algorithm}
\caption{ML Based 2-D Angle Estimation} \label{a2}
\begin{algorithmic}[1]
\STATE {Initialize $\mathbf{x}=[\hat{\phi}~\hat{\varphi}]^T$ (starting point), $M>1$, $\alpha\in(0,0.5)$, and $\beta\in(0,1)$.}
\STATE {Calculate $\triangledown l(\mathbf{x})$ in \eqref{16} and $\Vert\triangledown l(\mathbf{x})\Vert$.}
        \WHILE{$\Vert\triangledown l(\mathbf{x})\Vert>\epsilon$ and $\frac{\partial l}{\partial\varphi}\ne0$}
           \STATE {Initialize $t=t_0/M$ by using \eqref{18}, calculate $l(\mathbf{x})$ and $l(\mathbf{x}-t\triangledown l(\mathbf{x}))$.}
                \WHILE{$l(\mathbf{x}-t\triangledown l(\mathbf{x}))>l(\mathbf{x})-\alpha t\Vert\triangledown l(\mathbf{x})\Vert^2$}
                     \STATE {Set $t=\beta t$ and calculate $l(\mathbf{x}-t\triangledown l(\mathbf{x}))$.}
                \ENDWHILE
           \STATE {Set $\mathbf{x}=\mathbf{x}-t\triangledown l(\mathbf{x})$, calculate $\triangledown l(\mathbf{x})$ and $\Vert\triangledown l(\mathbf{x})\Vert$.}
       \ENDWHILE
\STATE {Calculate $\hat{\phi}=\hat{\phi}-2\pi\times\lfloor\frac{\hat{\phi}}{2\pi}\rfloor$ and output $\mathbf{x}$.}
\end{algorithmic}
\end{algorithm}

\section{Numerical Results}

In this section, simulation results are provided to show the effectiveness of the proposed method and all mean square error (MSE) results are averages of $1000$ independent runs. We set $\frac{R}{\lambda}=\frac{1}{4}$, $s(t)=\sqrt{P}$, $\sigma^2=1$, $M=10,\alpha=0.3,\beta=0.5$, and $\epsilon=0.03$.

\begin{figure}[t] \centering
\includegraphics[scale=0.6]{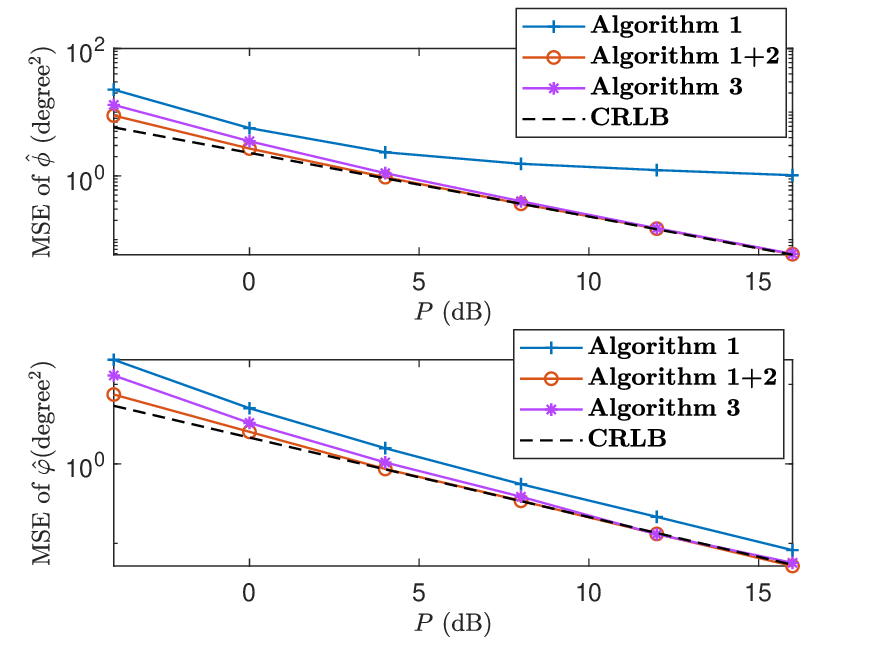}
\vspace{-0.2cm}\caption{MSE of angle estimation versus $P$ with $N=120$, $L=200$, $K=6$, $\delta=\{20,40,\ldots,120\}$, $\phi =110\degree$, and $\varphi = 44\degree$.}\vspace{0cm} \label{f2}
\end{figure}

Fig. \ref{f2} illustrates MSE results provided by \textbf{Algorithm \ref{a1}}, \textbf{Algorithm \ref{a2}}, and \textbf{Algorithm 3} proposed in \cite{Wu}. \textbf{Algorithm 3} also chooses the subset of array sensors $\delta$. It can be seen that due to quantization error, when $P\to\infty$, the MSE results of $\hat{\phi}$ provided by \textbf{Algorithm \ref{a1}} approach a positive constant number instead of zero. This indicates that \textbf{Algorithm \ref{a1}} cannot provide unbiased estimators. Also, it is very intuitive that the estimation performance of \textbf{Algorithm \ref{a1}} is inferior to that of \textbf{Algorithm 3}. However, the computational complexity of \textbf{Algorithm \ref{a1}} is to search $i^{\ast}$ in \eqref{10} and the operation of the inverse trigonometric function in \eqref{14}, which makes it computationally simpler than \textbf{Algorithm 3} and suitable for real-time signal processing.

%this search process only requires $\frac{N}{2}\times L$ complex multiplications, so \textbf{Algorithm \ref{a1}} can be utilized for a coarse angle estimation

We can see that when the estimates provided by \textbf{Algorithm \ref{a1}} are used as a starting point of \textbf{Algorithm \ref{a2}}, this joint algorithm can yield sufficient accuracy. Additionally, under low $P$ (e.g., $P=0~\mathrm{dB}$), the estimation performance of the joint algorithm is superior to that of \textbf{Algorithm 3} but both of them cannot achieve the Cramer-Rao lower bound (CRLB) \cite{Stoica2}. However, as $P$ increases (e.g., $P=12~\mathrm{dB}$), the joint algorithm and \textbf{Algorithm 3} almost achieve the CRLB. Fig. \ref{f2} demonstrates the effectiveness of the proposed method.

\begin{figure}[t] \centering
\includegraphics[scale=0.6]{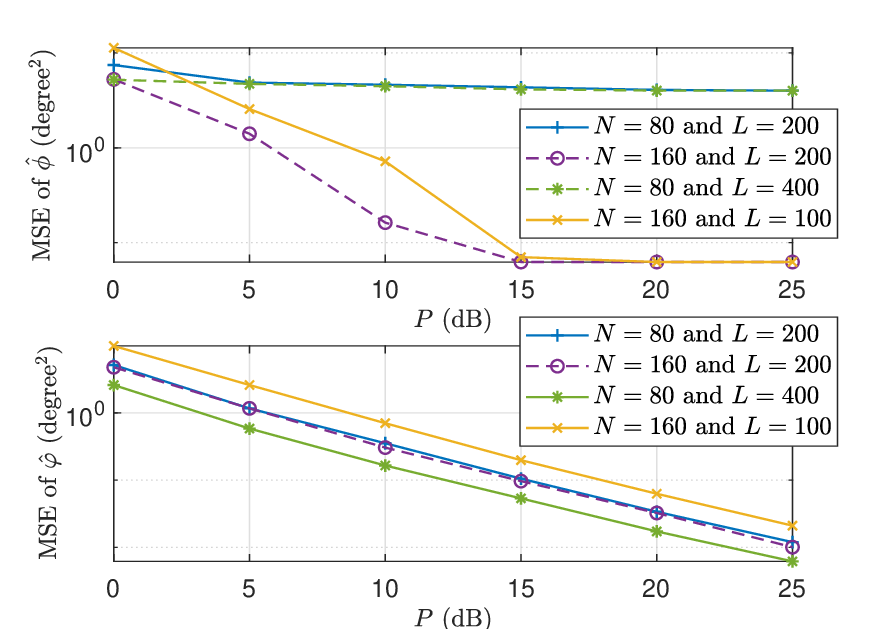}
\vspace{-0.2cm}\caption{MSE of angle estimation in \textbf{Algorithm 1} versus $P$ with $\phi=88\degree$ and $\varphi=44\degree$.}\vspace{0cm}\label{f3}
\end{figure}

To show the impacts of $N$ and $L$ on \textbf{Algorithm \ref{a1}}, Fig. \ref{f3} illustrates MSE results with different $N$ and $L$. We can see that when $P$ is small (e.g., $P=0~\mathrm{dB}$), the estimation performance of $\hat{\phi}$ with larger $L$ is better. The reason is that under low $P$, noise has a significant effect on covariances and increasing $L$ can mitigate the effect of noise. However, as $P$ increases, the estimation performance of $\hat{\phi}$ with larger $N$ becomes better. This is because under high $P$ (e.g., $P=20~\mathrm{dB}$), the effect of noise is negligible, larger $N$ (i.e., smaller quantization error) reduces MSE results. Similarly, under high $P$ (e.g., $P=25~\mathrm{dB}$), the estimation performances of $\hat{\phi}$ with identical $N$ but different $L$ become consistent. This is because the effects of noise and $L$ are not obvious, the consistent MSE results can be obtained due to the same quantization error.

Note that larger $L$ can improve the estimation performance of $\hat{\varphi}$ in spite of smaller $N$ and the estimation performances of $\hat{\varphi}$ with identical $L$ but different $N$ are almost consistent. This is because when $N$ is greater than a certain value, the quantization error of $\hat{\phi}$ has little effect on $\hat{\varphi}$. For example, assuming $\phi=\theta_{i^{\ast}}+\frac{\pi}{2}+\Delta\theta$, where $\Delta\theta$ is a quantization error and $-\frac{\pi}{N}\le\Delta\theta\le\frac{\pi}{N}$, \eqref{13a} can be rewritten as
\begin{eqnarray}     %16
\psi_{i^{\ast}+\frac{N}{4}} - \psi_{i^{\ast}+\frac{3}{4}N} = -2\zeta\sin(\varphi)\cos(\Delta\theta),\nonumber
\end{eqnarray}
where $\cos(\Delta\theta)\in [0.9992,1]$ if $N\ge80$. Hence, when $N\ge80$, increasing $N$ cannot efficiently improve the estimation performance of $\hat{\varphi}$ but increasing $L$ can reduce MSE results of $\hat{\varphi}$ efficiently.

\printbibliography % Output the bibliography

\end{document}